\documentclass[a4paper, 10pt]{jpconf}
\usepackage{graphicx}
\usepackage{amsmath}
\usepackage{amssymb}
\usepackage{iopams}
\usepackage{citesort}
\usepackage[utf8]{inputenc}
\usepackage[usenames,dvipsnames,svgnames]{xcolor}

\begin{document}
\title{Perturbations in a Chaplygin gas Cosmology}
\author{Shambel Sahlu$^{1,2,3}$, Heba Sami $^{3}$, Anna-Mia Swart$^{3}$, Thato Tsabone$^{3}$, Maye Elmardi$^{4}$ and Amare Abebe$^{3}$}
\address{$^{1}$ Astronomy and Astrophysics Department, Entoto Observatory and Research Center, 
~~~Ethiopian Space Science and Technology Institute, Addis Ababa, Ethiopia.}
\address{$^{2}$ Department of Physics, College of Natural and Computational Science, Wolkite University,
Wolkite, Ethiopia.}
\address{$^3$ Center for Space Research, North-West University, Mahikeng 2735, South Africa.}
\address{$^4$ Center for Space Research, North-West University, Potchefstroom 2520, South Africa.}

\ead{hebasami.abdulrahman@gmail.com}
\begin{abstract}
In this paper we study the perturbations of a cosmic multi-fluid medium consisting of radiation, dust and a Chaplygin gas. To do so, we follow the $1+3$ covariant formalism and derive the evolution equations of the fluctuations in the energy density for each species of fluid in the multi-fluid system. The solutions to these coupled systems of equations are then computed in both short-wavelength and long-wavelength modes. Our preliminary results suggest that unlike most dark energy models that discourage large-scale structure formation due to the rapid cosmological expansion (which gives little time for fluctuations to coalesce), the Chaplygin-gas model supports the formation of cosmic structures.  This is manifested in the solutions of the perturbation equations which show the growth of density fluctuations over time.
\end{abstract}
\section{Introduction} 
In the last two decades, it has become apparent that the standard cosmological model ($\Lambda$CDM) fails to explain the accelerating expansion of the Universe \cite{riess1998observational}, the existence of
dark matter in galaxy clusters \cite{fall1980formation}, the formation of the large-scale structure \cite{blumenthal1984formation}, the inherent inhomogeneity and anisotropy of the Universe on the small scales \cite{rees1980inhomogeneity} and so forth. To solve the first two problems, scholars  have proposed an exotic fluid, the so-called Chaplygin gas (CG), which acts as a cosmological fluid with an equation of state (EoS) of the form $p_{ch} = -A/\rho_{ch} ^\alpha$, where $p_{ch}$ and $\rho_{ch}$ are the pressure and energy density of the CG, and $A$ and $\alpha$ are constants such that $A>0$ and $0<\alpha\leq 1$ \cite{bento2002generalized}. From this EoS we can conclude that the EoS parameter for CG can be written as $w_{ch}=-A/{\rho_{ch}^{1+\alpha}}= -A/(A + B(1+z)^{3(1+\alpha)})$, where $z$ is the redshift. The purpose of this CG fluid is to substitute the  dark matter and dark-energy components of the Universe.  This model acts as dark-matter in the early universe and dark-energy in the late times of the cosmos. In a  universe in which we assume the energy density to consist of  matter and CG fluids, the conservation equation reads as
\small
\begin{equation}
\dot{\rho}_{t} + 3\frac{\dot{a}}{a}(\rho_{t}+p_{t}) = 0,
\end{equation}
\normalsize
 where $\rho_t$ is the energy density and $p_t$ is the pressure for the total fluid\footnote{When referring to matter, we mean either radiation or dust, and total fluid implies matter and Chaplygin gas.}, $a$ is the cosmological scale factor and  the subscript $t$ stands for total (matter+CG) fluids.  From this equation, the energy density becomes $\rho_{t}(z) =  \left[ A + {B}{(1+z)^{3(1+\alpha)}}\right]^{\frac{1}{1+\alpha}} + \rho_m (1+z)^{3(1+w_m)}$, where $B=e^{C(1+\alpha)}$ with $C$ a constant of integration.\footnote{$w_m$ is the equation of state parameter for the matter fluid which takes the value of $1/3$ for the radiation component ($w_r$) and $0$ for the dust component ($w_d$), and $w_t = p_t/\rho_t=p_t/(\rho_{ch}+\rho_m)$.} Whereas much has been said of the Chaplygin gas as an alternative to a unified dark matter and dark energy model that mimics the cosmic expansion history of a Friedmann-Lema\^{\i}tre-Robertson-Walker (FLRW) background,  to our knowledge there is no work in the literature that considered the cosmological perturbations of this fluid model in the 1+3 covariant formalism \cite{ellis1989covarianta,dunsby1991gauge,abebe2012covariant}. Henceforth we derive the density perturbation equations and present the solutions both analytically and numerically. From the results, we analyse cosmological implications as far as large-scale structure formation is concerned on both sub- and super-horizon scales \cite{abebe2012covariant}. The outline of the manuscript is as follows: in the next section we review the basic spatial gradient variables and we also derive the linear evolution equations. We apply scalar and harmonic decomposition  techniques and obtain the wave-number dependent energy density fluctuations for both CG and matter fluids. In Sec. \ref{numerical}, we present the  analytical and numerical solutions of density perturbations by considering radiation-CG, dust-CG and CG  like fluids for both wave-length ranges. We then devote Sec. \ref{discussions} to discussions of our results and the conclusions. 
{\section{Perturbations}
For a perfect-fluid system, the following fluid equations hold: $\dot{\rho}_t = -\Theta(\rho_t+p_t)\;,$ and $ (\rho_t+p_t)\dot{u}_a +\tilde{\nabla}_a p_t =0\;,$ from which one can conclude that the 4-acceleration is $\dot{u}_a = \alpha w_{ch} \rho_{ch} D^{ch}_a/(a(\rho_t + p_t))-w_m\rho_m D^{m}_a/(a(\rho_t + p_t))\;.$
 Another key equation for a general fluid is the so-called Raychaudhuri equation which can be expressed as $\dot{\Theta} = - \Theta^2 /3 - (\rho_t +3p_t)/2-\tilde{\nabla}^a\dot{u}_a\;,$ where $\Theta$ is the cosmological expansion scalar. After finding the comoving gradients of the cosmological expansion and the comoving fractional density gradient of the matter components in a covariant and gauge-invariant way, and taking the harmonically decomposed scalar parts, it can be shown that the scalar perturbations for the matter and CG energy densities, respectively, evolve according to
 \small
\begin{multline}
\ddot{\Delta}^k_m =  \Theta(w_{m}-\frac{2}{3})\dot{\Delta}_m+
\bigg\{(1+w_{m})\bigg[ \frac{1}{2}(1+3w_{m})+\frac{\Theta^2 w_{m} }{3\rho_t(1+w_{t})}
 + \frac{w_m(1 +3 w_t) }{2(1+ w_{t})}+ \dfrac{k^{2} w_{m}}{a^{2} \rho_{t}}\bigg]\rho_m  +\bigg[{\frac{1}{3}w_{m}\Theta^2}\\
 -\frac{w_{m}}{2}(1+w_t)\rho_t\bigg\} \Delta^k_m 
+(1+w_{m})
 \bigg[\frac{1}{2}-\frac{3w_{ch}}{2}-\frac{\Theta^2 w_{ch} }{3\rho_t(1 + w_{t})} - \frac{w_{ch}(1+ 3w_t ) }{2(1+w_t)}+ \dfrac{k^{2} w_{ch}}{a^{2} \rho_{t}}\bigg]\rho_{ch}~\Delta^k_{ch}\;,
\end{multline}
\begin{multline}
 \ddot{\Delta}^k_{ch} = -\Theta\left(w_{ch}+\frac{2}{3}\right)\dot{\Delta}^k_{ch}
 +(1+w_{ch})\bigg[ \frac{1}{2}(1+3w_m)+\frac{\Theta^2 w_m }{3\rho_t(1+w_{t})} + \frac{w_m(1 +3 w_t) }{2(1+ w_{t})}
 + \dfrac{k^{2} w_{m}}{a^{2} \rho_{t}}\bigg]\rho_m\Delta^k_m \\
  +\bigg\{(1+w_{ch})\bigg[\frac{1}{2}-\frac{3 w_{ch}}{2}-\frac{ \Theta^2 w_{ch} }{3\rho_t(1 + w_{t})} - \frac{w_{ch}(1+ 3w_t ) }{2(1+w_t)} 
 + \dfrac{k^{2} w_{ch}}{a^{2} \rho_{t}}\bigg]\rho_{ch}
-w_{ch}\left(\frac{1}{3}\Theta^2 + \frac{1}{2}(1 +3w_t)\rho_t\right)\\
 +\Theta^2w_{ch}\left[2(1+2w_{ch})\rho_{ch}+\frac{2}{3}\right]\bigg\}\Delta^k_{ch}\;,~~
\end{multline}
\normalsize
where $ k = 2\pi a/\lambda,  $ $k$ being the co-moving wave-number and $\lambda$ the wavelength of the perturbations. In GR, with normal forms of matter, one would obtain a closed second-order equation of the density fluctuations and the equations are generally easier to solve.  But here we get a coupled system of second-order equations for the density fluctuations of both matter and CG, the solutions to which are more complicated to compute analytically.  As such, we consider short-wavelength ($k/aH\gg1$) and  long-wavelength ($k/aH\ll1$) limits of the perturbations \cite{abebe2012covariant} and analyse large-scale structure implications of the resulting solutions.  

In redshift space, the corresponding equations can respectively be found to be \footnote{A single dot ( $\dot{}$ )  and double dots ( $\ddot{}$ ) represent the first- and second-order derivatives, respectively, with respect to time, $t$, and a single prime ( ${'}$ ) and double prime ( ${''}$ ) represent the first- and second-order derivatives, respectively, with respect to redshift, $z$.}}
\small
 \begin{multline}
{\Delta}''_m =\frac{1}{1+z}(\frac{1}{2}-4w_m){\Delta}'_m+
 \bigg\{\frac{(1+w_m)}{(1+z)^2}\bigg[ \frac{1}{2}(1+3w_m)+\frac{ w_m }{(\Omega_{ch}+\Omega_{m})(1+w_t)} +  \frac{w_m(1 +3 w_t) }{2(1+ w_t)}\\
  +\dfrac{9\pi^2(1+w_m)^2w_m}{3\lambda^2(1+z)^{3(1+w_m)} (\Omega_{ch}+\Omega_{m})}\bigg]3\Omega_m  +\bigg[{3w_m}-\frac{3}{2}w_m(\Omega_{ch}+\Omega_{m})(1+w_t)\bigg\} \Delta^k_m 
  +\frac{(1+w_m)}{(1+z)^2}\bigg[\frac{1}{2}\\-\frac{3\alpha w_{ch}}{2}-\frac{\alpha w_{ch} }{(\Omega_{ch}+\Omega_m)(1 + w_t)} - \frac{\alpha w_{ch}(1+ 3w_t ) }{2(1+w_t)}
  +\dfrac{9\pi^2(1+w_m)^2 \alpha w_{ch}}{3\lambda^2(1+z)^{3(1+w_m)}(\Omega_{ch}+\Omega_{m})}\bigg]3\Omega_{ch}\Delta^k_{ch}\;, \label{DMfluctuations}
\end{multline}
\begin{multline}
{\Delta}''_{ch} = \frac{3\sqrt{4}-6}{\sqrt{4}(1+z)}\left(w_{ch}(1+\alpha)-w_{ch}+\frac{2}{3}\right){\Delta}'_{ch}  +\frac{(1+w_{ch})}{(1+z)^2}\bigg[ \frac{1}{2}(1+3w_m)+ \frac{ w_m }{3(\Omega_m+\Omega_{ch})(1+w_t)} \\
  + \frac{w_m(1 +3 w_t) }{2(1+ w_t)}+ \dfrac{9\pi^2(1+w_m)^2w_m}{3\lambda^2(1+z)^{3(1+w_m)} (\Omega_{ch}+\Omega_{m})}\bigg]3\Omega_m\Delta^k_m +\frac{1}{(1+z)^2}\bigg\{{(1+w_{ch})}\bigg[\frac{1}{2}-\frac{3\alpha w_{ch}}{2} \\
  -\frac{\alpha  w_{ch} }{(\Omega_m+\Omega_{ch})(1 + w_t)} - \frac{\alpha w_{ch}(1+ 3w_t ) }{2(1+w_t)}+ \dfrac{9\pi^2(1+w_m)^2 \alpha w_{ch}}{\lambda^2(A+B(1+z)^{3(1+\alpha)})^\frac{1}{1+\alpha}(\Omega_{ch}+\Omega_{m})}\bigg]3\Omega_{ch}\\
  -w_{ch}\left(3 + \frac{3}{2}(1 +3w_t)(\Omega_m +\Omega_{ch})\right)+9w_{ch}\left[(1+\alpha)(1+2w_{ch})3\Omega_{ch}+\frac{2}{3}\right]\bigg\}\Delta^k_{ch} \;.\label{DMf1}
 \end{multline}
\normalsize
\section{Results and Discussion}\label{numerical}
{For further discussion of the growth of the energy density fluctuations with redshift, we assume non-interacting fluids, namely radiation-CG, dust-CG, and only CG fluids in the following subsections.\footnote{For all analysis, we consider the original CG model,  $\alpha =1$.} }
\subsection{GR and $\Lambda$CDM approaches}
{We first consider the well-known results of GR and $\Lambda$CDM approaches. For the case where the entire universe has been filled with dust and radiation fluids, the evolution equation becomes}
\small
 \begin{multline}\label{GRapproach1211}
 \frac{d^2\Delta_m(z)}{dz^2} -\frac{1}{2(1+z)}\frac{d\Delta_m(z)}{dz}
  -\frac{1}{(1+z)^2} \bigg[ \frac{3}{2} \Omega_m \left(1+3w_m\right)(1-w_m)+6w_m\Omega_{\Lambda} 
 -w_m{k^2}(1+z)^2\bigg]\Delta_m(z) = 0
\end{multline}
\normalsize
By eliminating the cosmological constant term, $\Lambda$, the $\Lambda$CDM model reduces to GR as shown in \cite{sahlu2020scalar}.
For illustrative purpose, we redefined the normalized energy density for fluid as $\delta(z)= {\Delta(z)}$ (as was done in \cite{sahlu2020scalar}), and  $\delta _{in} = \Delta(z_{in})$ is the initial value of the density contrast, $ \Delta_{}(z)$, at $z_{in} = 1100$.
\begin{figure}[h!]
\begin{minipage}{0.48\textwidth}
\includegraphics[width=0.9\textwidth ,height=4cm]{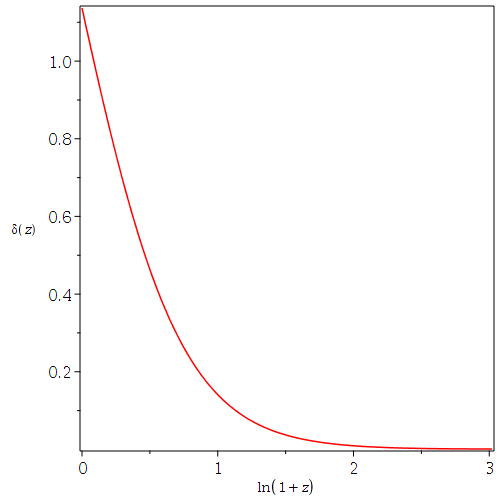}
 \caption{$\delta (z)$ versus $z$  for dust for GR.}
 \label{fig:dustGR}
\end{minipage}
\qquad
\begin{minipage}{0.48\textwidth}
\includegraphics[width=0.9\textwidth ,height=4cm]{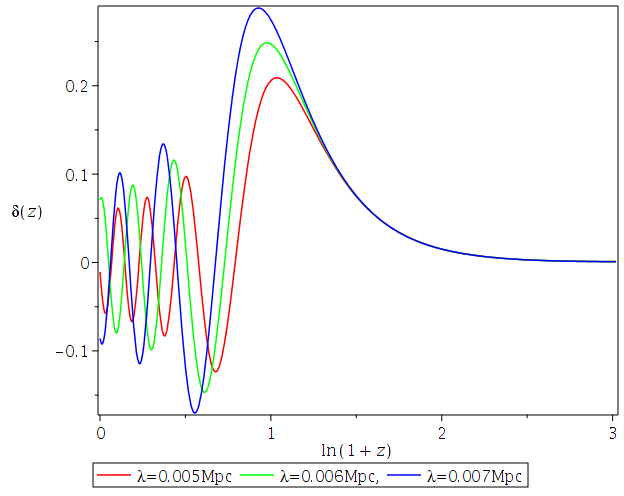}
 \caption{$\delta (z)$ versus $z$  for short wavelengths for $\Lambda$CDM with different wavelength.}
 \label{fig:shortGR}
\end{minipage}
\begin{minipage}{0.38\textwidth}
\includegraphics[width=\textwidth ,height=4cm]{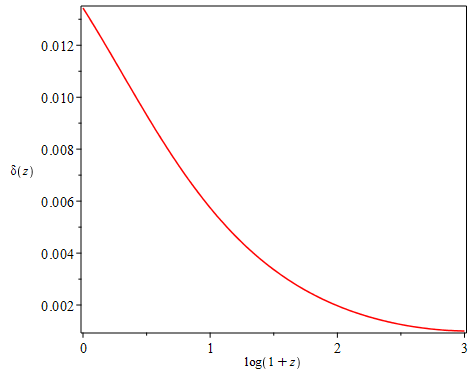}
\caption{ $\delta (z)$ versus $z$  for long wavelengths for $\Lambda$CDM.}
\label{fig:longGR}
\end{minipage} 
\qquad
\begin{minipage}{0.56\textwidth}
\vspace{0.5cm}
\small
 The numerical results for a dust dominated universe, and a radiation dominated universe are presented in Fig. \ref{fig:dustGR}, and  Fig. \ref{fig:shortGR} and \ref{fig:longGR},  for short and long-wavelength modes, respectively. From the plots, we observe the growth of matter density fluctuations with cosmic-time. For short-wavelength, the amplitude has an oscillatory behavior.  The variation of CMB temperature detected observationally is of the order of $10^{-3}$ \cite{smoot1992structure}. Therefore, we set the initial conditions at $\delta _{in}= \Delta _{i}(z_{in}\simeq 1100)= 10^{-3}$ and $\dot{\delta}_{in}=\dot{\Delta}_{i} (z_{in}=1100)= 0$.  The normalized energy densities are defined as $\Omega_m+\Omega_\Lambda = 1$.
 \normalsize
\end{minipage} 
\end{figure}
\subsection{Radiation-CG dominated Universe}
\begin{figure}[h!]
\begin{minipage}{0.48\textwidth}
\includegraphics[width=0.9\textwidth ,height=4cm]{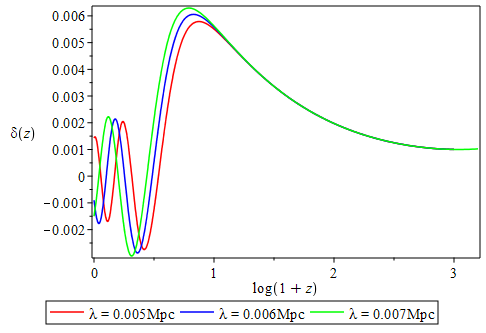}
 \caption{$\delta (z)$ versus $z$  for short wavelengths for radiation-CG fluids.}
 \label{fig:short}
\end{minipage} 
\qquad
\begin{minipage}{0.48\textwidth}
\includegraphics[width=0.9\textwidth ,height=4cm]{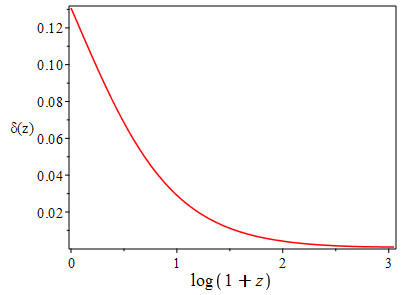}
\caption{ $\delta (z)$ versus $z$  for long wave-lengths for radiation-CG fluids.}
\label{fig:long}
\end{minipage} 
\end{figure}

The EoS parameter of the total fluid consisting of radiation (implying that $w_m=w_r$) and CG is given as
\small
 \begin{equation}
 w_t = \frac{\rho_{0,r}(1+z)^4-3\frac{A}{\sqrt{A+B(1+z)^{6}}}}{3\rho_{0,r}(1+z)^4 +3\sqrt{A+B(1+z)^{6}}}\;, 
 \end{equation}
 \normalsize
 where $\rho_{0,r}$ is the initial energy density of radiation at the time of the Big Bang. In the early universe, the redshift is large and CG acts as dark matter, whereas, in the late universe (where redshift is small and negative) the CG behaves like dark energy. The EoS parameter, $w_{t}$, approaches $-1$ when $z \ll 1$ (which means that the dominant component in late times is dark energy, corresponding to our $\Lambda$CDM model predictions).
 { Now, taking large redshift, $z\gg 1$, then $w_t \approx 1/3 = w_r$ (corresponding to a dominant radiation component in early times). 
If we assume  a dominant radiation fluid, and background CG fluid, we have $\Delta_r\gg\Delta_{ch}$, i.e., $\Delta_{ch}$ is very small. Similar analysis is done in \cite{abebe2012covariant} for a dust-radiation system.} Then the solutions of Eq. \eqref{DMfluctuations}, for the short-wavelength range, are found to be
\small
\begin{eqnarray}
\Delta(z) =&& C_1(1+z)^\frac{1}{12}\mbox{BesselJ}\bigg(\frac{\Sigma}{24(\Omega_{ch}+\Omega_r)},\frac{4\pi\sqrt{\Omega_r}}{3\lambda\sqrt{\Omega_{ch}+\Omega_r}(1+z)^2}\bigg)  \nonumber\\&&
+ C_2(1+z)^\frac{1}{12}\mbox{BesselY}\bigg(\frac{\Sigma}{24(\Omega_{ch}+\Omega_r)},\frac{4\pi\sqrt{\Omega_r}}{3\lambda\sqrt{\Omega_{ch}+\Omega_r}(1+z)^2}\bigg)\;,\label{short-solution}
\end{eqnarray}
\normalsize
where $\Sigma = \sqrt{624\Omega^2_r -96\Omega^2_{ch}+528\Omega_{ch}\Omega_r +145\Omega_{ch}+289\Omega_r (\Omega_{ch} + \Omega_r)}$, $C_1$ and $C_2$ are integration constants and BesselJ and BesselY are the first- and second-order Bessel functions. The solution of Eq. \eqref{DMfluctuations} for the long-wavelength range is given as
\small
\begin{eqnarray}
 \Delta(z) = C_3(1+z)^{\Omega_{ch} + \Omega_r + \frac{\psi}{{12(\Omega_{ch}+\Omega_r)}}} + C_4(1+z)^{-(\Omega_{ch} + \Omega_r - \frac{\psi}{{12(\Omega_{ch}+\Omega_r)}})}\;, \label{longsolutions}
\end{eqnarray}
\normalsize
where $\psi ={\sqrt{432\Omega^2_{ch}\Omega_r + 1152\Omega_{ch}\Omega^2_r - 96\Omega^3_{ch} +624\Omega^3_r+145\Omega^2_{ch}+289\Omega^2_r+434\Omega_{ch}\Omega_r}}\;.$
 We assume initial conditions given  as {{$\Delta_{in} \equiv \Delta^k (z_{in} = 1100) = 10^{-3}$}} and ${\Delta}'_{in} \equiv {\Delta}'(z_{in}=1100) =  0$  for every mode, $k$, to deal with the growth of matter fluctuations \cite{abebe2012covariant}. Therefore, we determine  the integration constant $C_i$ ($i$ = 1,2,3,4,...) by imposing these initial conditions. The numerical result of Eq. \eqref{short-solution} is presented in Fig. \ref{fig:short}. In this figure we observe the oscillatory motions of the density perturbations in the short-wavelength modes. The solutions for for the long-wavelength modes (given by Eq. \eqref{longsolutions}) are shown in  Fig. \ref{fig:long}. {In this figure, we clearly observe the growth of the density fluctuations.} We used $\Omega_r = 4.48\times 10^{-5}$ \cite{chavanis2015cosmology}.
  \subsection{Dust-CG dominated Universe}
\begin{figure}[h!]
 \begin{minipage}{0.48\textwidth}
\includegraphics[width=0.9\textwidth ,height=4cm]{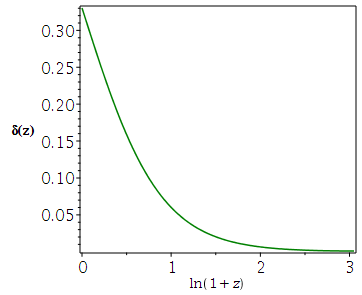}
 \caption{$\delta (z)$ versus $z$  for dust-CG fluids  and for $\Omega_d = 1-\Omega_{ch}$. We used $\Omega_d = 0.32$ \cite{ade2016planck}.}
 \label{fig:dust-cg}
\end{minipage}
\qquad
\begin{minipage}{0.48\textwidth}
\includegraphics[width=0.9\textwidth ,height=4cm]{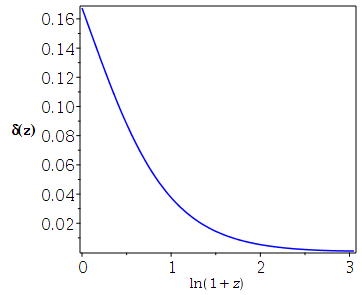}
\caption{ $\delta (z)$ versus $z$  for CG alone, using $1-\Omega_r -\Omega_d = \Omega_{ch}$.}
\label{fig:Chaplygin}
\end{minipage} 
\end{figure}

If we consider a total fluid which consists of dust (implying that $w_m=w_d$) and CG, then the EoS parameter of the total fluid becomes
\small
\begin{equation}
w_t = \frac{- \frac{A}{\sqrt{A+B(1+z)^{6}}}}{\rho_{0,d}(1+z)^3 +\sqrt{A+B(1+z)^{6}}}.
\end{equation}
\normalsize
For large values of redshift, $z \gg 1$ (i.e. in the early universe), $w_t \approx 0$ (corresponding to a dominant dust component in early times). When $z\ll 1$, the EoS parameter becomes $w_t \approx  -1$, which again corresponds to dark energy being the dominant component at late times. The solutions of our  evolution equations for the density perturbations, which are given by Eq. \eqref{DMfluctuations}, in a dust-CG system is presented graphically in Fig. \ref{fig:dust-cg} and reads
\small
  \begin{eqnarray}
   \Delta(z) = C_5(1+z)^{\frac{3}{4}+\frac{1}{4}\sqrt{9+24\Omega_d}} +C_6(1+z)^{\frac{3}{4}-\frac{1}{4}\sqrt{9+24\Omega_d}}\;.
  \end{eqnarray}
  \normalsize
\subsection{CG-dominated Universe}
{When the energy density of the CG fluid is dominant over that of the matter fluid,} we can let $\Delta_{ch}\gg\Delta_{m}$, causing $\Delta_m$ to be very small. The solution of Eq. \eqref{DMf1} then reads as
\small
\begin{eqnarray}
\Delta(z) = C_7\log(1+z)\sin(\Omega_r +\Omega_d +\Omega_{ch} +5) +C_8\log(1+z)\cos(\Omega_r +\Omega_d +\Omega_{ch} +5)\;.
\end{eqnarray}
\normalsize
The numerical results are presented in Fig. \ref{fig:Chaplygin} and depicts the growth of the CG density fluctuations in terms of redshift, and the role of CG fluid for the formation of large-scale structure.
{In this paper we observe that CG acts as a dark energy and it is an alternative approach to explaining the current phase of universe. From all numerical results, we observe that the CG component becomes dominant over time and the growth of matter density contrast is fast growing with cosmic time compared to the matter-CG fluid system, since the Universe becomes less structured in the accelerating universe. }  
We think our current results show that even at the level of the perturbations, the CG fluid offers an alternative to the narrative of large-scale structure formation. This is because, contrary to what one would expect in a dark-energy-dominated universe where there would be less chance of large-scale structure formation due to the rapid cosmological expansion, we see the growth of the density perturbations with time.
\section{Conclusions}\label{discussions}
In this work, we explored the solutions of cosmological perturbations in a multi-fluid cosmic medium where one of the fluids is a Chaplygin gas. We applied the $1+3$ covariant and gauge-invariant formalism to define the spatial gradient variables and applied scalar and harmonic decomposition methods to analyse the scalar perturbations of  the different energy densities involved. We considered different systems such as radiation-CG, dust-CG and CG fluids in both short- and long-wavelength modes to present the numerical and analytical solutions to the perturbation equations. Our results show that, at least in the simplest CG model, the formation of large-scale structures is enhanced, rather than discouraged (as one would expect from dark energy fluid models), since all our preliminary calculations show the growth of density fluctuations with time. 
\small
 \section*{Acknowledgments}
 SS gratefully acknowledges financial support from WKU, EORC and ESSTI, as well as the hospitality of the Physics Department of North-West University (NWU) during the preparation of part of this manuscript. HS acknowledges the financial support from the Mwalimu Nyerere  African Union scholarship and the National Research Foundation (NRF) free-standing scholarship.  TT has been funded through the National Research Foundation (NRF) free-standing scholarship.   AS acknowledges funding through the National Astrophysics and Space Science Program (NASSP) scholarship. ME acknowledges that this work is supported by an NWU/NRF postdoctoral fellowship. AA acknowledges that this work is based on the research supported in part by the NRF of South Africa with grant numbers 109257 and 112131.
\providecommand{\newblock}{}

\end{document}